\documentclass[onecolumn,preprintnumbers,amsmath,amssymb,superscriptaddress,nofootinbib,12pt]{revtex4}

\usepackage{epsf}
\usepackage{graphicx}  
\usepackage{dcolumn}   
\usepackage{bm}        

\begin{document}

\preprint{KU-TP 039}

\title{Black Holes in Gravity with Conformal Anomaly and Logarithmic Term
in Black Hole Entropy}

\author{Rong-Gen Cai}\email{cairg@itp.ac.cn}
\address{
Key Laboratory of Frontiers in Theoretical Physics, Institute of
Theoretical Physics, Chinese Academy of Sciences, P.O. Box 2735,
Beijing 100190, China}
\address{
Department of Physics, Kinki University, Higashi-Osaka, Osaka
577-8502, Japan}

\author{Li-Ming Cao}\email{caolm@itp.ac.cn}
\address{
Department of Physics, Kinki University, Higashi-Osaka, Osaka
577-8502, Japan}

\author{Nobuyoshi Ohta}\email{ohtan@phys.kindai.ac.jp}
\address{
Department of Physics, Kinki University, Higashi-Osaka, Osaka
577-8502, Japan}

\vspace*{2.cm}
\begin{abstract}
We present a class of exact analytic and static, spherically
symmetric black hole solutions in the semi-classical Einstein
equations with Weyl anomaly. The solutions have two branches, one
is asymptotically flat and the other asymptotically de Sitter. We
study thermodynamic properties of the black hole solutions and
find that there exists a logarithmic correction to the well-known
Bekenstein-Hawking area entropy.  The logarithmic term might come
from non-local terms in the effective action of gravity theories.
The appearance of the logarithmic term in the gravity side is
quite important in the sense that with this term one is able to
compare black hole entropy up to the subleading order, in the
gravity side and in  the microscopic statistical interpretation
side.

\end{abstract}
\maketitle

\newpage

Conformal (Weyl) anomalies have a long history~\cite{Duff1} (For a
nice review see \cite{Duff2}). The conformal anomaly is not only a
quite important concept in quantum field theory in curved spaces,
but also has a variety of applications in cosmology, black hole
physics, string theory and statistical mechanics. For example,
Christensen and Fulling showed that in two spacetime dimensions,
the Hawking radiation has a very close relation to the trace
anomaly~\cite{CF}. Recently this idea has been employed to study
Hawking radiation of higher dimensional black holes~\cite{RW}. In
the aspect of application of conformal anomaly in cosmology, a
well-known result is that the anomaly can lead to a significant
inflation model~\cite{Star,Hawking,Nojiri}. The trace anomaly
might also be related to the well-known cosmological constant
problem~\cite{Duff2}.

In a Friedmann-Robertson-Walker (FRW) spacetime setting, the
trace anomaly can completely determine  its corresponding
energy-momentum tensor due to the fact that the spacetime is
homogeneous and isotropic. In a black hole spacetime background,
the trace anomaly can also completely determine the
energy-momentum tensor  in the two-dimensional case;
 in the four-dimensional static, spherically symmetric black
hole case, however, it determines the energy-momentum tensor up to
an arbitrary function of position~\cite{CF}, which causes the
difficulty in studying the backreaction of the trace anomaly to the
black hole spacetime~\cite{AMV}.

On the other hand, in quantum field theory in curved spaces, it is
required to include the backreaction of the quantum fields to the
spacetime geometry itself~\cite{BD}
\begin{equation}
\label{eq1} R_{ab}-\frac{1}{2}R g_{ab} = 8\pi G \langle T_{ab}
\rangle,
\end{equation}
where $\langle T_{ab} \rangle$ is the effective energy-momentum
tensor by quantum loops. However, the expression of the effective
energy-momentum tensor is usually quite complicated so that in
general one is not able to find exact analytic solutions to the
semi-classical Einstein equations (\ref{eq1}). Therefore, one
usually considers perturbative solutions to equations (\ref{eq1})
in the literature: one first assumes some background solutions to
the vacuum Einstein equations, then calculates the vacuum
expectation value of the effective energy-momentum tensor $\langle
T_{ab} \rangle$ in the fixed background, and finally discusses
 the backreaction effect of the quantum fields perturbatively.

In this paper we report on some exact solutions of the
semi-classical Einstein equations (\ref{eq1}) with conformal
anomaly. To our knowledge, it is the first time to obtain exactly
nontrivial black hole solutions to the Einstein equations with
conformal anomaly. In addition, we find that the entropy of the
black hole solutions has a logarithmic term, in addition to the
well-known horizon area term. The appearance of the logarithmic
term is quite interesting. According to the statistical
interpretation of black hole entropy in some quantum theories of
gravity such as string theory and loop quantum gravity, on a very
general ground, it can be argued that usually the leading term of
statistical degrees of freedom gives the Bekenstein-Hawking
horizon area term, while the subleading term is a logarithmic
term. In the gravity side, according to Wald' entropy formula of
black holes~\cite{Wald}, however, it is quite difficult to produce
such a logarithmic term in the black hole entropy in some
effective local theory of gravity even with higher derivative
curvature terms. It means that they do not get matched to the
entropy in the subleading order in the gravity side and in the
microscopic statistical interpretation side. Here the conformal
anomaly corrected black hole solutions provide a possible
interpretation for the appearance of the logarithmic term in the
field theory side.

Let us start our discussions on our theory. In even dimensions, quantum
effects will destroy the conformal invariance of classical
conformal field theories.  One loop quantum corrections lead to a
trace anomaly of the energy-momentum tensor of conformal field
theories. In general, the trace anomaly has the
form~\cite{Duff2,Deser}
\begin{equation}
\label{eq2}
 g^{ab}\langle T_{ab}\rangle= \lambda I_{(2n)} - \alpha
E_{(2n)}\, ,
\end{equation}
where $\lambda$ and $\alpha$ are two positive constants depending
on the degrees of freedom of quantum fields and their values will
not affect our discussions.  The term $I_{(2n)}$  is a polynomial
of Weyl tensor, which is called type B anomaly, while $E_{(2n)}$
is Euler characteristic of  $2n$-dimensional spacetime, named type
A anomaly in~\cite{Deser}.  In four dimensions, $I_{(4)}$ is given
by
\begin{equation}
I_{(4)}=C_{abcd}C^{abcd}\, ,
\end{equation}
and $E_{(4)}$ is just Gauss-Bonnet term
\begin{equation}
E_{(4)}=R^2-4 R_{ab}R^{ab}+R_{abcd}R^{abcd}\, .
\end{equation}
We now look for exact solutions to equations of motion (\ref{eq1})
with corresponding energy-momentum tensor to the trace anomaly
(\ref{eq2}). For simplicity, consider a static, spherically
symmetric spacetime of the form
\begin{equation}
\label{eq5} g_{ab}dx^adx^b=-f(r)dt^2+\frac{1}{g(r)}dr^2+ r^2
d\Omega_{2}^2\, ,
\end{equation}
where $f(r)$ and $g(r)$ are two functions of the radial coordinate
$r$ only, and $d\Omega_2^2$ is the line element of a
two-dimensional sphere with unit radius. As for the vacuum
expectation value $\langle T_{ab}\rangle$ of effective
energy-momentum tensor, we only know the following two constraint
conditions: (i) its trace is given by (\ref{eq2}), and (ii) it
must be covariantly conserved, namely, $\nabla_a \langle
T^{ab}\rangle =0$. As was known in \cite{CF}, in that case, one is
not able to determine completely the energy-momentum tensor. To go
on, we have to impose another condition to determine the
corresponding energy-momentum tensor to the trace anomaly. Note
that to have a nontrivial black hole solution with metric
(\ref{eq5}), at a black hole horizon, say $r_+$, the
energy-momentum tensor must satisfy~\cite{CJ}: $\langle T^t_{\
t}(r)\rangle|_{r=r_+} =\langle T^r_{\ r}(r)\rangle|_{r=r_+}$.
However, this condition does not fix the energy-momentum tensor
which is necessary to find exact solutions.
Henceforth we make a further assumption that this relation
holds not only at the black hole horizon, but also in the whole
spacetime, namely,
\begin{equation}
\label{eq6}
\langle T^t_{\ t}(r)\rangle =\langle T^r_{\ r}(r)\rangle.
\end{equation}
Though at this point this is merely an assumption to find exact
solutions, there are certain examples that satisfy this relation,
and we will discuss this further towards the end of this paper.

With the symmetry of the metric (\ref{eq5}), we can define
\begin{equation}
\label{eq7}
 \langle T^t_{\ t}\rangle =-\rho(r)\, ,\qquad \langle
T^r_{\ r}\rangle =p(r)\, ,\qquad \langle T^{\theta}_{\
\theta}\rangle =\langle T^{\phi}_{\ \phi}\rangle =p_{\bot}(r)\, .
\end{equation}
 With the help of  assumption (\ref{eq6}), one can easily show
 that  $f(r)=g(r)$. On the other hand, the constraints
(i) and (ii) turn out to be
\begin{equation}
\label{eq8} -\rho+p+2 p_{\bot}=-\alpha E_{(4)}\, ,
\end{equation}
and
\begin{equation}
\label{eq9}
 4f(p-p_{\bot})+(\rho+p)rf'+2 r p'f=0\, ,
\end{equation}
where we have set $\lambda=0$ and a prime denotes the derivative
with respect to $r$ and $ E_{(4)}=
\frac{2}{r^2}\left((1-f)^2\right)''$.  Now put $p_{\bot}$ from
(\ref{eq9})  and $\rho=-p$ from (\ref{eq6}) into (\ref{eq8}), one
has
\begin{equation}
rp'+4p + \alpha E_{(4)}=0\, ,
\end{equation}
which has the solution
\begin{equation}
\label{eq11}
 p=\frac{2\alpha}{r^4}(1-f)(1-f+2rf')-\frac{q}{r^4}\, ,
\end{equation}
where $q$ is an integration constant. We can then easily obtain the
energy density $\rho =-p$ and the transverse pressure
$p_{\bot}=-\alpha E_{(4)}/2 -2p$. Substituting these results
together with (\ref{eq7}) into the semi-classical Einstein equations
(\ref{eq1}), we obtain a solution with metric function
\begin{equation}
\label{eq12}
 f (r)=1- \frac{r^2}{4\tilde \alpha}\left(1\pm
\sqrt{1-\frac{16\tilde \alpha G M}{r^3}+\frac{8 \tilde \alpha
Q}{r^4}}\right)\, ,
\end{equation}
where $M$ is an integration constant,  $\tilde \alpha = 8\pi G\alpha
$ and $Q= 8\pi G q$. We have checked indeed the solution (\ref{eq5})
with $g(r)=f(r)$ satisfies all components of the semi-classical
Einstein equations (\ref{eq1}). The solution has two branches with
$``\pm" $ in (\ref{eq12}), respectively. That is, there exist two
vacuum solutions when $M=Q=0$: one is the Minkowski spacetime for
the branch with $``-"$ sign, and the other is a de Sitter space with
the effective radius $\ell_{\rm eff}=\sqrt{2\tilde \alpha}$ for the
branch with $``+"$ sign. The solution (\ref{eq5}) with (\ref{eq12})
looks very like the Gauss-Bonnet black hole solution presented
in~\cite{BD2}, but here our solution is four dimensional. In fact,
our solution looks more like the black hole solution with a negative
constant curvature space in the Gauss-Bonnet gravity~\cite{DM}. But
again, those solutions presented in~\cite{DM} describe higher
dimensional ($D \ge 6$) objects.

In the large $r$ limit, for the branch with $``-"$ sign,  we have
\begin{equation}
f(r)\approx 1- \frac{2G M}{r}+\frac{Q}{r^2} +{\cal O}(r^{-4})\, .
\end{equation}
We can clearly see that the solution behaves like the
Reissner-Norstr\"om (RN) solution if $Q>0$ and that the
integration constant $M$ is nothing but the mass of the solution.
The meaning of the integration constant $Q$ will be discussed
later. On the other hand, for the branch with $``+"$ sign, the
behavior of large $r$ looks like
\begin{equation}
f(r)\approx 1- \frac{r^2}{\ell_{\rm eff}^2} +\frac{2G
M}{r}-\frac{Q}{r^2} +{\cal O}(r^{-4})\,.
\end{equation}
This is the behavior of a Reissner-Nordstr\"om-de Sitter solution
if $M<0$ and $Q<0$.

It is expected that the trace anomaly will not change the vacuum
of the theories, therefore we believe that the branch with $``+"$
is unstable like the black hole solutions in the Gauss-Bonnet
gravity~\cite{BD2}. In addition, in the cosmology setup, the de
Sitter solution is found unstable~\cite{Star}. Therefore we will
restrict ourselves to the branch with $``-"$ sign in what follows.

Now we turn to study the meaning of the integration constant $Q$.
Taking the limit $\tilde \alpha \to 0$, one has from the solution
(\ref{eq12}) that
\begin{equation}
\label{eq15}
 f(r) =1- \frac{2G M}{r}+\frac{Q}{r^2}.
\end{equation}
Clearly the vacuum solution to the Einstein equations (\ref{eq1})
without the trace anomaly term must be the Schwarzschild solution
in the metric form (\ref{eq5}). This means one should have $Q=0$.
Does it mean that one has to always take $Q=0$ when the trace anomaly
term appears. The answer is negative. In fact, we can see from the
solution (\ref{eq11}) that the term associated with the
integration constant $q$ is nothing but a ``dark radiation" term
with $\rho_d =p_{\bot}=-p =q/r^4$. Such a term satisfies the two
constraints (i) and (ii), and is also consistent with the symmetry
of the metric (\ref{eq5}). Therefore with the two constraints (i)
and (ii), and the assumption (\ref{eq6}), one cannot
exclude the existence of the ``dark radiation".  In addition, we
will see that when a Maxwell field is present, the electric charge
square $Q_e^2$ will appear in the same place as $Q$ in the
solution (\ref{eq12}) [see the solution (\ref{eq22})] . Therefore
the integration constant $Q$ corresponds to a quantity to be explained as
$U(1)$ conserved charge square of some conformal field theory.
Thus we keep this term $Q$ and have $Q>0$ in order to have a
positive energy density associated with this term.

Next let us have a look at the singularity of the solution. The
square of Riemann tensor is given by
\begin{equation}
R_{abcd}R^{abcd}\sim \frac{2}{\tilde
\alpha}\left(\frac{Q}{r^4}-\frac{2GM}{r^3}\right)+\cdots \,,
\end{equation}
in the small $r$ limit. There is therefore a singularity at the
origin of $r$ as the RN solution in general relativity. Besides,
there exists another potential singularity where the square root
vanishes in (\ref{eq12}). It is quite interesting to compare the
small $r$ behavior of the square of Riemann tensor between the
solution (\ref{eq12}) and the usual RN solution, which has the
expansion behavior
\begin{equation}
R_{abcd}R^{abcd}\sim \frac{56 Q^2}{r^8}-\frac{96GMQ}{r^7}+
\frac{48 G^2 M^2}{r^6}+\cdots \, .
\end{equation}
We can clearly see that the backreaction of the Weyl anomaly
drastically softens the singularity at the origin.

Now we turn to thermodynamic properties of the black hole solution
(\ref{eq12}). The black hole horizon satisfies $f(r)=0$,
which has two roots $ r_{\pm}=GM\pm \sqrt{G^2M^2 -(Q-2\tilde
\alpha)}$. This implies that the black hole could have two
horizons, a degenerated horizon and naked singularity if
$G^2M^2>Q-2\tilde \alpha$, $G^2M^2=Q-2\tilde \alpha$ and
$G^2M^2<Q-2\tilde \alpha$, respectively. We assume the existence
of the black hole horizon, and then the ADM mass of the black hole can
be expressed as
\begin{equation}
GM=\frac{r_+}{2}+\frac{Q}{2r_+}-\frac{\tilde \alpha}{r_+}\, .
\end{equation}
The Hawking temperature is easy to give by calculating surface
gravity at the horizon
\begin{equation}
\label{eq19}
 T=\frac{1}{4\pi}f'(r_+) =\frac{r_+}{4\pi(r_+^2-4\tilde
\alpha)}\left(1-\frac{Q}{r_+^2}+\frac{2\tilde \alpha}{r_+^2}\right).
\end{equation}
We can see from the solution (\ref{eq12}) that  in order to have a
horizon, the horizon radius must satisfy $r_+^2> 4\tilde \alpha$.
Therefore the behavior of the Hawking temperature of the black
hole is quite similar to the one for the usual RN black hole: the
temperature starts from zero for an extremal black hole,
monotonically increases and reaches its maximum at some horizon
radius, and then monotonically decreases forever as the horizon
radius is increased.

Black hole entropy is an important quantity in black hole
thermodynamics. Due to the appearance of the Weyl anomaly term,
the well-known area formula must no longer hold. Usually Wald's
entropy formula is a powerful tool to calculate black hole
entropy~\cite{Wald}. Unfortunately Wald's formula cannot be used
either here since we do not know the effective action of the
anomaly term. Instead we try to obtain the black hole entropy by
employing the first law of black hole thermodynamics, $dM
=TdS+\dots$, where $\dots$ stands for some work terms, if any.
Integrating the first law yields
\begin{equation}
S=\int T^{-1}\left(\frac{\partial M}{\partial
r_+}\right)_{Q}dr_+=\frac{\pi r_+^2}{G} - \frac{4 \pi \tilde
\alpha}{G} \ln r_+^2 + S_0\, .
\end{equation}
Here $S_0$ is an integration constant, which unfortunately we
cannot fix  because of the existence of the logarithmic term. The
entropy can also be expressed as
\begin{equation}
\label{eq21}
 S= \frac{A}{4G}- \frac{4\pi\tilde \alpha}{G} \ln
\frac{A}{A_0}\,,
\end{equation}
where $A=4\pi r_+^2$ is the horizon area and $A_0$ is a constant
with dimension of area. Clearly the first term is just the
well-known Bekenstein-Hawking area term of black hole entropy,
while the appearance of the logarithmic term in (\ref{eq21}) is
worth saying a few words.

First, we notice that our entropy formula (\ref{eq21}) is completely
the same as the one associated with the apparent horizon of an FRW
universe with the anomaly term~\cite{Lidsey}, which comes out by
investigating the relation between the modified Friedmann equation
and the first law of thermodynamics~\cite{CCH}. This partially
supports the conjecture that the entropy formula associated with the
apparent horizon of an FRW universe is the same as the one of black
hole horizon in the same gravity theory~\cite{ACK}. Second, as we
mentioned above, a logarithmic term is universally present as a
subleading correction to the Bekenstein-Hawking area entropy, in the
microscopic statistical interpretation of black hole entropy such as
in string theory, loop quantum gravity, thermal and/or quantum
fluctuations in a fixed black hole background (see references cited
in \cite{CCH}). However, in the gravity side, it is quite difficult
(if not impossible) to produce such a term from a local effective
gravity theory, based on Wald's entropy formula~\cite{Wald}, which
says that black hole entropy is a Noether charge, given by an
integral of the variation of the Lagrangian of the effective gravity
theory with respect to Riemann tensor over the spatial cross section
of black hole horizon. Thus this is a serious challenge to match the
subleading term of black hole entropy in both sides.  Here it seems
the first time to give the logarithmic term from the gravity
side~\footnote{We notice that such logarithmic term also appears in
the entropy of black holes of Ho\v{r}ava-Lifshitz
gravity~\cite{CCO}. The Ho\v{r}ava-Lifshitz theory is a
nonrelativistic gravity theory, and therefore a non-local gravity
theory.}. Such a logarithmic term coming from the non-local trace
anomaly may shed some lights for the origin in the gravity side.
Indeed, we notice that in the paper~\cite{Solo}, Solodukhin argued
on scaling and dimensionality ground that such a logarithmic
correction to a four-dimensional Schwarzschild black hole entropy
could come from rather complicated non-local functionals in the
low-energy effective action of string theory and the coefficient of
the logarithmic term is proportional to the four-dimensional central
charge which comes from the integrated conformal anomaly for the
zero-mass fields in the theory.  Therefore the logarithmic term in
our case has seemingly the same origin as that discussed
in~\cite{Solo}. However, we here obtain such a term by directly
solving the semi-classical Einstein equations (\ref{eq1}) without
any approximation.

Some remarks are in order. (i) Note that we have set $\lambda=0$ in
order to find the exact analytic solution (\ref{eq12}). Namely we
have only considered the type A anomaly~\cite{Deser}. At the moment
it is not clear whether an analytic solution can be derived once the
type B anomaly (the Weyl tensor square term) is included. In
particular it involves higher derivative terms. But we believe that
with the same assumption, one is able to find some exact solutions
at least numerically in that case. (ii) In this paper we only
discussed the vacuum solution of the semi-classical Einstein
equations without any classical fields. If the cosmological constant
$\Lambda$ and some gauge fields are present, in the static,
spherically symmetric case, we can easily find the exact solution as
\begin{equation}
\label{eq22}
f(r) =g(r)=1- \frac{r^2}{4\tilde \alpha}\left(1\pm
\sqrt{1+\frac{4\tilde \alpha}{\ell^2}-\frac{16\tilde \alpha G
M}{r^3}+\frac{8 \tilde \alpha (Q+Q_e^2)}{r^4}}\right)\, ,
\end{equation}
where $Q_e$ is the electric charge of Maxwell field and
$\ell^2=3\Lambda$. (iii) In higher even ($2n(n>2)$) dimensions, the
type A anomaly will be given by Euler Characteristic of
$2n$-dimensional spacetime. With the same approach, one is able to
get exactly analytic solutions. In that case, a logarithmic term
will also appear in the black hole entropy. (iv) In the spirit of
AdS/CFT correspondence, a conformal field theory occurs in a brane
world. Hence generalizing our solution to the brane world scenarios
is of great interest. These issues are currently under
investigation. (v) The another remark is concerned with the
assumption (\ref{eq6}). First, let us note that the condition that
$\langle T^t_t\rangle|_{r+r_+} =\langle T^r_r \rangle |_{r=r_+}$
holds at a black hole horizon is a requirement for a regular
horizon~\cite{CJ}, and it is satisfied by the stress-energy tensor
of the trace anomaly in Schwarzschild black hole and
Reissner-Nordstr\"om black hole backgrounds (see, for example,
\cite{AMV}). Second, Eq.~(\ref{eq6}) is just an assumption in this
paper. However, this is indeed satisfied by some conformal field
theories. One typical example is the Maxwell field: for the electric
field produced by an electric charge $q$ at the origin $r=0$, its
stress-energy tensor has the form $T^t_t =T^r_r \sim -q^2/r^4$.
Furthermore, the stress-energy tensor for a non-linear extension of
Maxwell field also satisfies the assumption in the background
(\ref{eq5}). In addition, this assumption holds as well even for the
effective energy-momentum tensor of higher derivative terms such as
Gauss-Bonnet term and more general Lovelock terms~\cite{BD2}. The
nonlcal nature of the anomaly term gives complicated stress-energy
tensor in general, but we could consider the special situation in
which this relation is effectively satisfied. It is true that in
this case our solution (\ref{eq12}) is not a general static
spherically symmetric solution, but represents a limited kind of
black hole solutions with backreaction of trace anomaly. The
consistent entropy formula (\ref{eq21}) with the one in the
cosmological setting signifies that our solution (\ref{eq12})
captures some features of the backreaction of trace anomaly.
Needless to say, it is of great interest to find exact analytic
solutions without the assumption (\ref{eq6}). (vi) Finally we
mention that when the temperature (\ref{eq19}) vanishes, the
solution (\ref{eq5}) with (\ref{eq12}) stands for an extremal black
hole, whose near horizon geometry is $AdS_2 \times S^2$. Indeed it
is shown in \cite{Sol2} that $AdS_2\times S^2$ is an exact solution
of the semi-classical Einstein equations (\ref{eq1}) in presence of
the Maxwell field. Thus our result is completely consistent with
that of \cite{Sol2}.

In summary we have presented a class of exact analytic black hole
solutions in the semi-classical Einstein equations with Weyl
anomaly. The set of solutions is parameterized by two integration
constants, one is the mass of the solutions and the other is a
$U(1)$ conserved charge of some classical conformal field theory.
The solutions have two branches, one is asymptotically flat and
the other asymptotically de Sitter. We have argued that the branch
with asymptotic de Sitter behavior is unstable.  We have studied
thermodynamic properties of the black hole solutions and found
that there exists a logarithmic correction to the well-known
Bekenstein-Hawking area entropy. We have discussed the
implications of the logarithmic term. It might come from non-local
terms in the effective action of gravity theories. The appearance
of the logarithmic term in black hole entropy implies that a full
quantum theory of gravity must be a non-local theory.

\section*{Acknowledgements}
RGC thanks S.P. Kim, K. Maeda, S. Mukohyama, B. Wang and Z.Y. Zhu
for helpful discussions for some relevant topics, and Kinki
University for warm hospitality.  RGC is supported partially by
grants from NSFC, China (No. 10821504 and No. 10975168) and a grant
from MSTC, China (No. 2010CB833004). LMC is supported by JSPS
fellowship No. P 09225. NO is supported in part by the Grant-in-Aid
for Scientific Research Fund of the JSPS No. 20540283, and also by
the Japan-U.K. Research Cooperative Program.

\end{document}